# From Cosmic Explosions to Terrestrial Fires?: A Reply

Adrian L. Melott[a] and Brian C. Thomas[b]

a Department of Physics and Astronomy, University of Kansas, Lawrence, Kansas 66045 USA.  melott@ku.edu

b Department of Physics and Astronomy, Washburn University, Topeka, Kansas 66621, USA.  brian.thomas@washburn.edu

ABSTRACT

Deschamps and Mottez (hereafter DM) argue that the Gauss-Matuyama terrestrial magnetic field reversal may have left a vanishing main dipole moment to the field for a time of order 10,000 years. They say this may have allowed an enhanced cosmic ray flux, boosting the effect we proposed in Melott and Thomas (2019). We point out that the bulk of the cosmic ray flux from a nearby supernova should be too energetic, up to a million times more energetic than the limits of deflection by the terrestrial magnetic field. In fact, only those highly energetic ones will directly reach the troposphere, relevant for cloud-to-ground lightning.

From Cosmic Explosions to Terrestrial Fires?: A Reply A.L. Melott and B.C. Thomas. J. Geology 128, online ahead of print.
https://www.journals.uchicago.edu/doi/abs/10.1086/7097501

From cosmic explosions to terrestrial fires? (A.L. Melott and B.C. Thomas) Journal of Geology, 127, 475-481. DOI: 10.1086/703418 (2019)

1. Introduction

DM argue that the Gauss-Matuyama magnetic field reversal at the Pliocene-Pleistocene boundary would have greatly lowered the overall ability of the Earth's magnetic field for of order $10^4$ years, allowing many more cosmic rays to reach the Earth's atmosphere and surface. It is argued that this would enhance the effect we proposed earlier.

This argument is valid for most of the cosmic rays we received today, which are below a billion ($10^9$) eV (1 GeV) in energy. These cosmic rays mostly affect only the upper atmosphere directly. However, as shown in Figures 2 and 3 of our paper, the input from nearby supernova have energies up to a million times greater. Earth's magnetic field is only effective for particles with rigidity below about 20 GV (Smart and Shea 2009); for protons this roughly 20 GeV, well below the energies to which our modeled supernova cosmic rays extend. This is the basis of their ability to penetrate to the troposphere and ionize it, and the basis of our hypothesis on the great increase of cloud-to-ground lightning. But since such cosmic rays are almost not at all deflected by the Earth's magnetic field, they would therefore not be measurably enhanced by a magnetic field reversal.

Ionization of the stratosphere and ozone depletion is possible by the cosmic rays that are deflected by the terrestrial magnetic field, and might be slightly enhanced by the reversal. But this was not the topic of our paper.

**References Cited**

From Cosmic Explosions to Terrestrial Fires?: A Discussion. F. Deschamps and F. Mottez. J. Geology 128, online ahead of print.
https://www.journals.uchicago.edu/doi/abs/10.1086/709750